%% file: main.tex
\documentclass[twocolumn]{aastex631}
\usepackage{amsmath, amsfonts}
\usepackage{acronym}
\usepackage{graphicx}
\usepackage{subfigure}

\graphicspath{./figures}

\begin{document}

\title{Structure and Skewness of the Effective Inspiral Spin Distribution of Binary Black Hole Mergers}


\author[0000-0001-7852-7484]{Sharan Banagiri}
\email{sharan.banagiri@monash.edu}
\affiliation{Center for Interdisciplinary Exploration and Research in Astrophysics (CIERA), Northwestern University, 1800 Sherman Ave, Evanston, IL 60201, USA}
\affiliation{School of Physics and Astronomy, Monash University, VIC 3800, Australia}
\affiliation{OzGrav: The ARC Centre of Excellence for Gravitational Wave Discovery, Clayton, VIC 3800, Australia}

\author[0000-0001-9892-177X]{Thomas A. Callister}
\affiliation{Kavli Institute for Cosmological Physics, The University of Chicago, Chicago, IL 60637, USA}

\author[/0000-0001-5525-6255]{Christian Adamcewicz}
\affiliation{School of Physics and Astronomy, Monash University, VIC 3800, Australia}
\affiliation{OzGrav: The ARC Centre of Excellence for Gravitational Wave Discovery, Clayton, VIC 3800, Australia}

\author[0000-0002-2077-4914]{Zoheyr Doctor}
\affiliation{Center for Interdisciplinary Exploration and Research in Astrophysics (CIERA), Northwestern University, 1800 Sherman Ave, Evanston, IL 60201, USA}
\affiliation{Department of Physics and Astronomy, Northwestern University, 2145 Sheridan Road, Evanston, IL 60208, USA}

\author[0000-0001-9236-5469]{Vicky\,Kalogera}
\affiliation{Center for Interdisciplinary Exploration and Research in Astrophysics (CIERA), Northwestern University, 1800 Sherman Ave, Evanston, IL 60201, USA}
\affiliation{Department of Physics and Astronomy, Northwestern University, 2145 Sheridan Road, Evanston, IL 60208, USA}
\affiliation{NSF-Simons AI Institute for the Sky (SkAI),172 E. Chestnut St., Chicago, IL 60611, USA}

\begin{abstract}
\noindent
The detection of gravitational waves has brought to light a population of binary black holes that merge within a Hubble time. Multiple formation channels can contribute to this population, making it difficult to definitively associate particular population features with underlying stellar physics. Black hole spins are considered an important discriminator between various channels, but they are less well-measured than masses, making conclusive astrophysical statements using spins difficult thus far. In this paper, we consider the distribution of the effective inspiral spin $\chi_{\rm eff}$ -- a quantity much better measured than individual component spins. We show that non-Gaussian features like skewness, asymmetry about zero, and multimodality can naturally arise in the $\chi_{\rm eff}$ distribution when multiple channels contribute to the population. Searching for such features, we find signs of skewness and asymmetry already in the current catalogs, but no statistically significant signs of bimodality. These features provide robust evidence for the presence of a subpopulation with spins preferentially aligned to the binary's orbital angular momentum; and we conservatively estimate the fraction of this subpopulation to be at least $12 \% - 17\%$ (at $90\%$ credibility). Our models do not find {a sharp excess} of non-spinning systems and instead find that at least $\sim 20 \%$ of the binaries have some degree of negative $\chi_{\rm eff}$. The data also suggest that, if preferentially aligned mergers form a significant fraction of the population, they must have small spins.
\end{abstract}


\acrodef{LVK}[LVK]{LIGO-Virgo-KAGRA}
\acrodef{BH}[BH]{black hole}
\acrodef{BBH}[BBH]{binary black hole}
\acrodef{BNS}[BNS]{binary neutron star}
\acrodef{CCSN}[CCSN]{core collapse supernova}
\acrodef{GW}[GW]{gravitational-wave}
\acrodef{LIGO}[LIGO]{Laser Interferometer Gravitational-Wave Observatory}
\acrodef{AGN}[AGN]{active galactic nuclei}
\acrodef{PPD}[PPD]{posterior population distribution}
\acrodef{PN}[PN]{post-Newtonian}

\section{Introduction}
\label{sec:intro}

The first detection of a \ac{BBH} merger by the \ac{LIGO}~\citep{ALIGO:2014pky} detectors happened in 2015~\citep{LIGOScientific:2016aoc}. By the end of their third observing run (O3) in 2020, the \ac{LVK}~\citep{ AVIRGO:2014yos, KAGRA:2020tym} collaborations have produced a catalog of $\sim 100$ gravitational-wave detection candidates~\citep{Catalog:GWTC3}, 
{the majority of which are \ac{BBH} mergers}. These detections have provided a glimpse into the population of extra-galactic stellar-mass \acp{BH} in binaries. As remnants of massive stellar evolution, it is expected that the population of \acp{BBH} contains signatures of the physics of stellar evolution.

As of the end of O3, we now understand broad details about \ac{BBH} mass, spin, and redshift distributions and have constrained the local \ac{BBH} merger rate to $\sim 10 - 100 \,{\rm Gpc}^{-3} {\rm yr}^{-1}$~\citep{RnP:GWTC2, RnP:GWTC3}. The merger rate has been established to very likely increase with redshift. The mass distribution of \ac{BBH} mergers falls roughly as a power law with overdensities at $\sim 10 M_{\odot}$ and $\sim 33 M_{\odot}$ and a preference for equal-mass pairing~\citep{RnP:GWTC3, Farah:2023swu}. The \ac{BH} spins are small compared to the measured spins of Galactic X-ray binaries~\citep{RnP:GWTC2, RnP:GWTC3, Corral-Santana:2015fud, Reynolds:2020jwt, Fishbach:2021xqi}. However, the connection between these broad population features and stellar astrophysics is not clear.

The physics of binary stellar evolution leading to \ac{BBH} formation is complex. Several key uncertainties remain in our understanding of stellar winds, mass transfer, common envelope evolution, and supernovae; see \cite{Mapelli:2018uds} and \cite{Mandel:2021smh} for reviews. The physics of natal kicks are also poorly understood, and observations suffer from systematics since almost all detected stellar-mass \acp{BH} are in binaries. Modeling stellar physics with high fidelity in population synthesis codes is computationally intensive, and as a result, codes often make simplifying assumptions about stellar evolution or use fits developed from isolated stellar evolution~\citep{Eldridge:2017, Giacobbo:2018, Breivik:2020, Riley:2022, Spera:2022, Fragos:2023, Andrews:2024}

Further complicating this is the possibility that multiple formation channels contribute to the expected population of \ac{BBH} mergers. These include isolated binary evolution ~\citep[e.g.,][]{Belczynski:2015tba, Giacobbo:2018etu, Stevenson:2017tfq,vandenHeuvel:2017pwp, Neijssel:2019irh, deMink:2016vkw, Marchant:2016wow, Bavera:2020uch}, dynamical channels such as globular clusters and nuclear star clusters~\citep[e.g.][]{Banerjee:2009, Antonini:2016gqe, Kremer:2019iul, Gerosa:2021mno, Mapelli:2020xeq}, dynamical triples~\citep[e.g.][]{Silsbee:2016djf, Naoz:2016, Rodriguez:2018jqu, Martinez:2020}, and \ac{AGN} disks~\citep[e.g.][]{Mckernan:2017ssq, Secunda:2018kar, Tagawa:2019osr, Sedda:2023big}. Uncertainties abound in the population synthesis predictions of these various channels. And while there are large differences between the different codes, the rates from the various channels broadly remain compatible with measurements from the \ac{LVK}~\citep{Mandel:2021smh}.

Spin tilts~\citep{Kalogera:1999tq, Rodriguez:2016vmx, Stevenson:2017dlk, Gerosa:2018wbw, Vitale:2015tea} are considered potential discriminators between formation channels. Several studies have attempted to check for signs of different formation channels by developing phenomenological mixture models~\citep[e.g.][]{Talbot:2017yur, Baibhav:2022qxm, Vitale:2022dpa, Pierra:2024fbl, Li:2023yyt} for spins and spin tilts of individual compact objects (henceforth component spins and spin tilts) in a binary. {While there are indications in the data of predominantly hierarchical contributions for mergers with masses greater than $40 M_{\odot}$~\citep{Pierra:2024fbl, Li:2023yyt, Antonini2024}, broad and clear constraints of the contributions of} multiple channels remain lacking. Component spins and spin tilts are much harder to measure than masses, as these impact the waveform at higher \ac{PN} orders. This, coupled with our uncertainties about the physics of the channels, has made it challenging to constrain their branching fractions.

On the other hand, there exists a combination of spins that is much better measured than component spin and spin tilts and is a constant of motion at 2\ac{PN} order~\citep{Gerosa:2015tea}. This is the so-called effective inspiral spin parameter $\chi_{\rm eff}$, which is the mass-weighted sum of the component spins projected along the orbital angular momentum~\citep{Damour:2001tu,Racine2008,Santamaria2010,Ajith2011},

\begin{equation}
\chi_{\rm eff} = \frac{\left(m_1 \vec{a_1} + m_2 \vec{a_2} \right) \cdot \hat{L} }{m_1 + m_2}.
\end{equation}
Here, $m_1$ and $m_2$ are the masses of the two components in the binary, $\vec{a}_1$ and $\vec{a}_2$ are the dimensionless spin vectors, and $\hat{L}$ is the unit vector along the direction of the orbital angular momentum.

Some analyses in the literature have modeled and directly fitted the $\chi_{\rm eff}$ distribution of \ac{BBH} mergers. \cite{Farr:2017uvj} was an early attempt to use $\chi_{\rm eff}$ to distinguish between different \ac{BBH} populations, but the study only considered certain representative population distributions without doing a full hierarchical Bayesian analysis. \cite{Roulet2019} and \cite{Miller:2020zox} introduced a truncated normal model that, when applied to data from the first three observing runs of the \ac{LVK} yields a mean at about $\chi_{\rm eff} \simeq 0.06$. The model also had support for a non-zero fraction of binaries having a small but negative $\chi_{\rm eff}$~\citep{RnP:GWTC3}. The better precision with which $\chi_{\rm eff}$ is measured also makes it useful for drawing astrophysical conclusions. \cite{Fishbach:2022lzq} used the lack of support for moderately large negative values of $\chi_{\rm eff}$ to limit the fraction of hierarchical mergers, i.e., mergers of \acp{BH} that are themselves products of earlier mergers. Finally,~\cite{Antonini2024} identified evidence that the effective spin distribution undergoes a transition at $m_1 \simeq 45\,M_\odot$, with more massive systems exhibiting a preferentially symmetric spin distribution.

Another fruitful line of inquiry has been the correlation of $\chi_{\rm eff}$ with various other \ac{BBH} parameters. Using data from up to the first half of the O3 run, \cite{Callister:2021fpo} found an anti-correlation between $\chi_{\rm eff}$ and $q$, the mass ratio of the \ac{BBH} mergers. This correlation was further tested and confirmed by \cite{Adamcewicz:2022hce} using the method of statistical copulas, and by \cite{RnP:GWTC3} in data up until the end of O3. \cite{Biscoveanu:2022qac} reported a correlation between the standard deviation of the $\chi_{\rm eff}$ distribution and redshift. The source of these correlations is not yet clearly understood, and it is unknown if they arise from a single formation channel or the interplay between multiple channels.

Given the current state of the field and our inability to yet identify any particularly dominant \ac{BBH} formation channel, we ask in this paper what signatures the $\chi_{\rm eff}$ distribution might exhibit in the presence of multiple channels. We broadly consider two formation channel types based on how the component spins are oriented: one that can capture a subpopulation where spins are randomly oriented and the other in which there is some structure to the tilt distribution. We show that using just the $\chi_{\rm eff}$ distribution, we can put conservative lower limits on sub-populations that can arise from diverse formation channels.

In Sec.~\ref{Sec:eff_spin_shape}, we consider the shape of the $\chi_{\rm eff}$ distribution and potential non-Gaussian signatures when we have contributions from different channel types. In Sec.~\ref{Sec:skew_eff_spin}, we focus on detecting skewness and asymmetry of the $\chi_{\rm eff}$ distribution using \ac{LVK} data, while in Sec.~\ref{Sec:mixture_models} we consider mixture models that more directly separate out the contribution of the two broad channel types. In Sec.~\ref{Sec:discussion}, we discuss these results and place them in the larger context of other studies of the $\chi_{\rm eff}$ distribution and its correlations with other parameters. Unless specified otherwise, all statistical statements in the paper are at $90\%$ credible levels. Throughout the paper, we use the convention that label `1' (`2') refers to the more-massive (less-massive) compact object in the binary.

\section{The shape of the effective inspiral spin distribution of BBH mergers}
\label{Sec:eff_spin_shape}

The proposition that component spin tilts, $\theta_1$ and $\theta_2$, can be used to learn about the formation channels of \ac{BBH} mergers depends, qualitatively, on the different channel types giving us different distributions of tilts. We can broadly divide formation channels into two categories.

\begin{figure}[t]

 \centering

 \includegraphics[width=0.48 \textwidth]{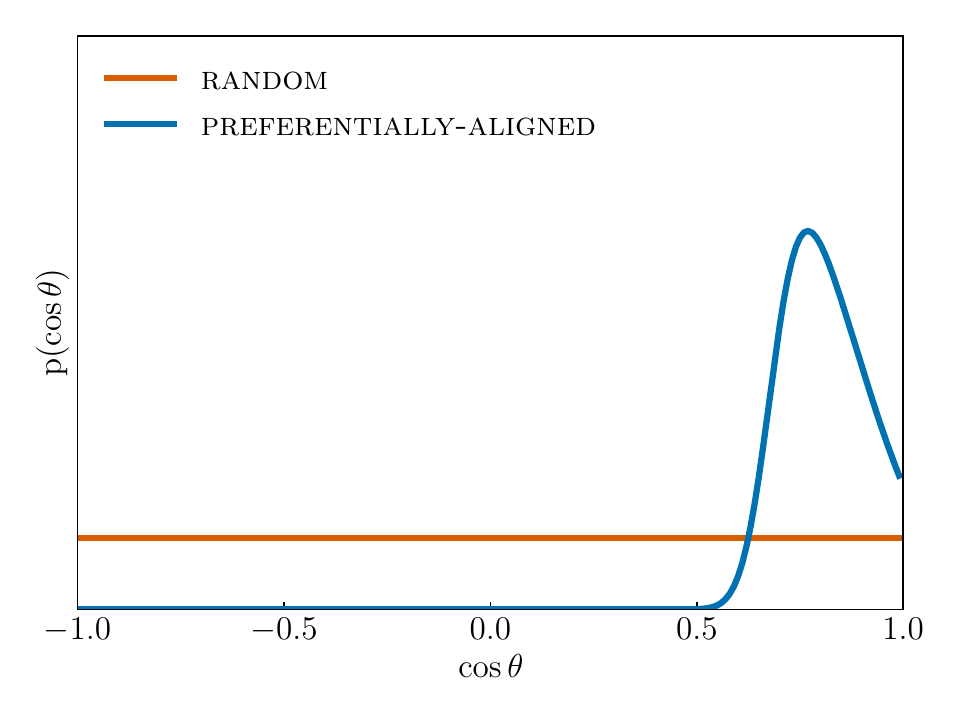}

\caption{A illustration of tilt angles under random and preferentially-aligned spin channels. The tilt distribution is uniform for a random spin channel. A preferentially-aligned spin channel can present some degree of alignment or anti-alignment. Furthermore, based on the physics of the preferentially-aligned spin channel, the tilt distribution can also be skewed about its mode.}

\label{fig:tilt-angles}

\end{figure}

\begin{figure*}[ht]

 \centering

\includegraphics[width=0.8\textwidth]{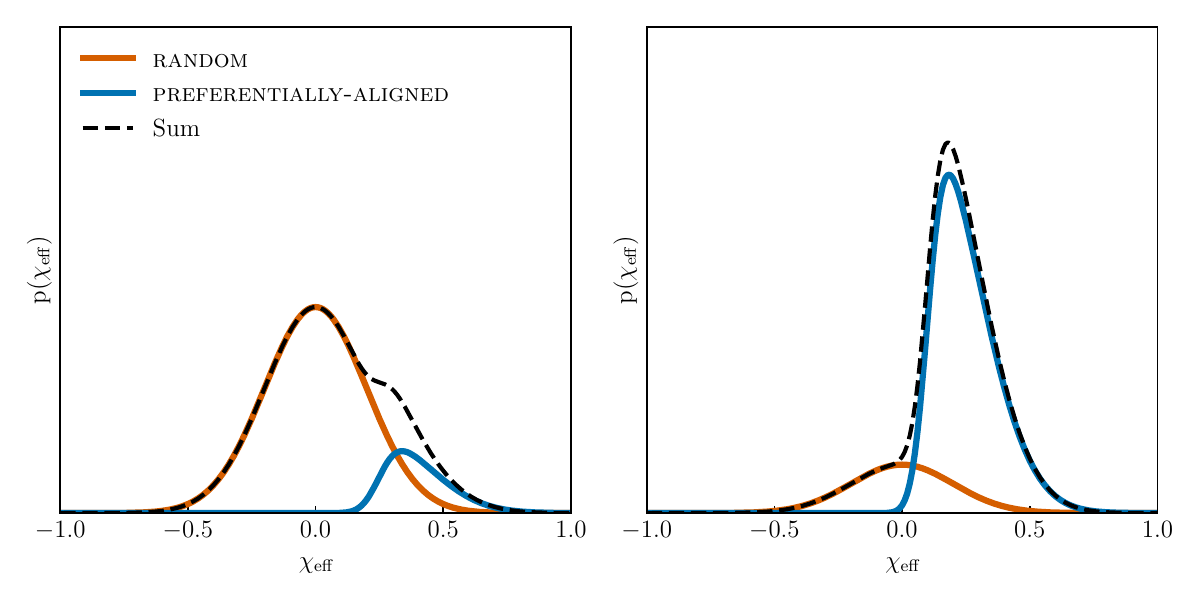}

 \caption{Illustrative plots showing how the effective distributions of a random and aligned population can look like and what the combined population can look like. The plot on the left shows the case where a random spin channel dominates the population, while the plot on the right illustrates a case where a positively preferentially-aligned spin channel dominates. }

 \label{fig:eff_spin_illustration}

\end{figure*}

\begin{enumerate}

 \item Random spin channels: In these channels, the \ac{BH} spin tilts are randomly aligned to the orbital angular momentum~(e.g., \cite{Rodriguez:2016vmx}). The prime examples of such channels are cluster formation channels, including globular and nuclear star clusters.

\item Preferentially-aligned spin channels: In these channels, the spins of the \acp{BH} have some (a priori unknown) degree of alignment or anti-alignment with the orbital angular momentum. Isolated binary evolution is the quintessential example of such a channel~\citep{Kalogera:1999tq, Gerosa:2018wbw}, but it has been suggested that \ac{BBH} mergers generated through the \ac{AGN} disk channel should also have some degree of preferential alignment~\citep{McKernan:2021nwk, Wang:2021yjf}.

\end{enumerate}

Figure~\ref{fig:tilt-angles} shows an illustrative example of the spin tilt-angle distribution for these two channel types. We note that, despite being a common assumption in empirical modeling~\citep[e.g.][]{Talbot:2017yur, RnP:GWTC3}, the spin tilt distribution for a preferentially-aligned spin channel need not be centered on perfect alignment~\citep{Vitale:2022dpa}. In the case of isolated binary evolution, the resultant \ac{BH} spins are dependend on several poorly understood mechanisms such as natal kicks, tides, and mass loss, and the interplay between these factors during the evolutionary history of the progenitor stars~\citep{Baibhav:2024}. Similar uncertainties are even higher in channels like the \ac{AGN} disks, where predictions hinge on modeling of complex gas physics~\citep[e.g.][]{McKernan:2019beu, Tagawa:2020dxe, Secunda:2020mhd}. Therefore, to be conservative, in this paper, we try not to make strong assumptions about the tilt distribution and thereby the $\chi_{\rm eff}$ distribution of preferentially-aligned channels.

Let us now consider what broad shapes the effective spin distributions might take for these two channel types. For a random spin channel such as cluster formation, we can safely assume that the probability of alignment and anti-alignment of a \ac{BH} is the same. Consequently, the $\chi_{\rm eff}$ distribution of a random spin channel will be symmetric about zero~\footnote{However, see \cite{Kiroglu:2025bbp} which posits that collisions between \acp{BBH} and individual stars in clusters can produce a population with small aligned spins.}, although the actual width and shape of the distribution will depend on the spin magnitude and mass distributions~(e.g., \cite{Antonini2024}). On the other hand, the $\chi_{\rm eff}$ distribution of the preferentially-aligned spin channel can peak away from zero and have some degree of asymmetry about its mode. The actual location of the peak and the shape of the distribution will depend on uncertain binary physics (or things like gas physics in AGNs) and the actual mass and spin magnitude distributions.

Fig.~\ref{fig:eff_spin_illustration} illustrates some potential shapes of the effective inspiral spin distribution when we combine a random and a preferentially-aligned spin channel. The figure on the left depicts the case where the random spin channel dominates, while the right shows the case where an aligned population dominates. While the figures themselves are purely illustrative, it helps us think of signatures we can expect for the $\chi_{\rm eff}$ distribution of the population in the presence of both aligned and random spin channel types.

\begin{enumerate}
\item Skewness: The joint distribution can potentially be skewed about the global mode of the distribution. This skewness can arise from the combination of preferentially-aligned and random spin channels or could predominantly reflect just the skewness of the former~\citep{Gerosa:2018wbw}.

\item Asymmetry about $\chi_{\rm eff} = 0$: The joint distribution can potentially be asymmetric about $\chi_{\rm eff} = 0$. Once again, this can arise from the combination of preferentially-aligned and random spin channels or just because the former dominates the population. As described earlier, a purely random spin channel will be entirely symmetric about zero~\citep{Vitale:2015tea,Rodriguez:2016vmx}.

\item Multimodality: In cases where both channels show a non-negligible contribution, the net $\chi_{\rm eff}$ distribution can be multimodal.
\end{enumerate}

In general, we expect that multiple channels might impart signatures of non-Gaussianity into the $\chi_{\rm eff}$ distribution. Driven by these observations, in Secs.~\ref{Sec:skew_eff_spin} and \ref{Sec:mixture_models}, we fit empirical models to detect these features in \ac{GW} data from the \ac{LVK}, starting with skewness and asymmetry in the next section. For the rest of this paper, we use asymmetry to refer to asymmetry about $\chi_{\rm eff} = 0$ unless specified otherwise.

\begin{figure*}[ht]
\subfigure[PPD for the \textsc{skewnormal} model]{
\centering
\includegraphics[width=0.48\textwidth]{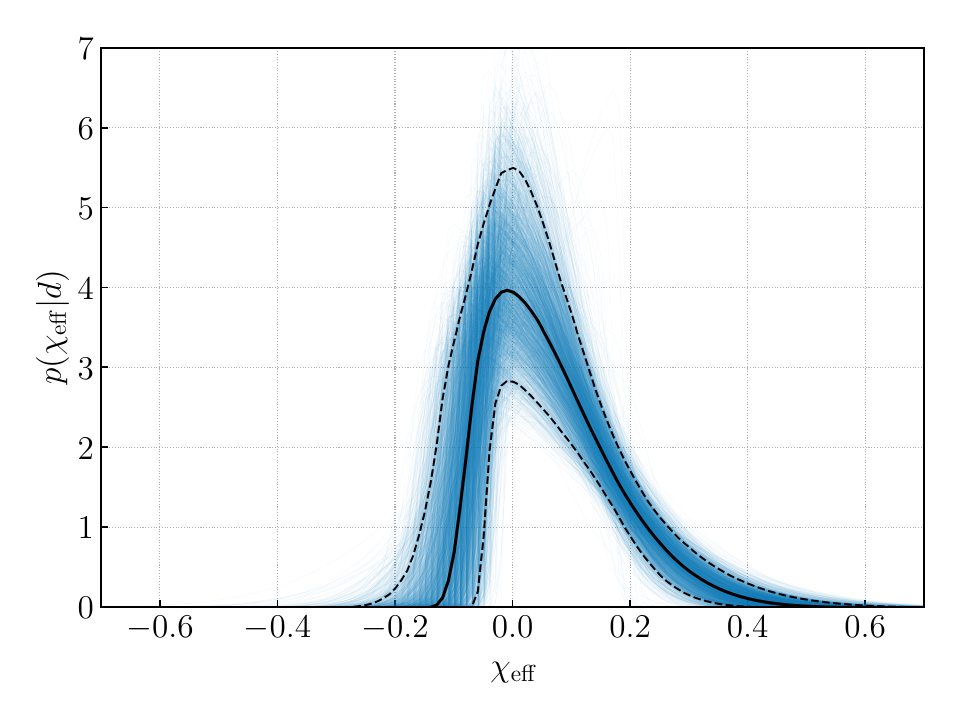} 
\label{Fig:skewnorm_ppd}
}
\subfigure[PPD for the $\varepsilon$-\textsc{skewnormal} model]{
\centering
\includegraphics[width=0.48\textwidth]{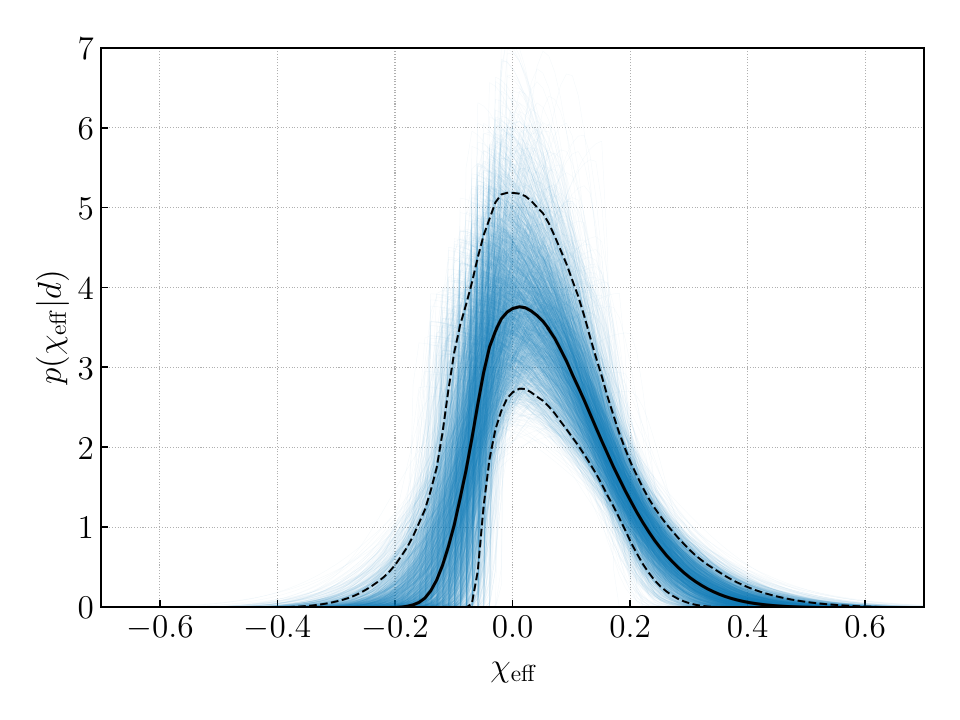}
\label{Fig:eps_skewnorm_ppd}
}
\caption{The $\chi_{\rm eff}$ \ac{PPD} under the \textsc{skewnormal} model on the left, and the $\varepsilon$-\textsc{skewnormal} model on the right, using data from the first three observing runs of the \ac{LVK}. The blue traces show the probability distribution for individual draws from the posterior. The solid black line is the median value of $p(\chi_{\rm eff} | d)$ at each $\chi_{\rm eff}$ while the dashed black lines are the $90 \%$ credible levels.}
 \label{Fig:ppd}
\end{figure*}

\section{Skewness and Asymmetry}
\label{Sec:skew_eff_spin}

First, we check for signs of skewness and asymmetry in the BBH $\chi_{\rm eff}$ distribution, which can be interpreted as signs of an aligned sub-population. Indeed, there is already some evidence in the literature that both of these features exist. As described in Sec.~\ref{sec:intro}, the Gaussian model fit to $\chi_{\rm eff}$ already exhibits an asymmetry about zero~\citep{Miller:2020zox, RnP:GWTC3}. Moreover, when accounting for the $\chi_{\rm eff} - q$ correlation, the marginal distribution in $\chi_{\rm eff}$ also shows a positive skewness~\citep{Callister:2021fpo, Adamcewicz:2022hce}. We return to the astrophysical implications of skewness and the $\chi_{\rm eff}- q$ correlation in Sec.~\ref{sec:chi_eff_q_corr}.

To directly capture a skewed and asymmetric distribution in $\chi_{\rm eff}$, we introduce the \textsc{skewnormal} model~\citep{Azzalini:1985c}, which modifies the normal distribution to allow each of the tails to have different shapes. The probability density function is given by:

\begin{equation}
\small
 \begin{split}
 {\rm SN} \left( \chi_{\rm eff} | \mu_{\rm eff}, \sigma_{\rm eff}^2 , \eta_{\rm eff} \right) \propto \, & \mathcal{N} \left( \chi_{\rm eff} | \mu_{\rm eff}, \sigma_{\rm eff}^2 \right) \\ \times &\left[ 1 + {\rm erf} \left( \eta_{\rm eff} \frac{\chi_{\rm eff} - \mu_{\rm eff}, }{\sigma_{\rm eff} \sqrt{2}}\right) \right], 
\end{split}
\label{Eq:skewnormal}
\end{equation}
where $ \mu_{\rm eff}$ and $\sigma_{\rm eff}$ are the location and the scale parameters and $\eta_{\rm eff}$ is the skewness parameter. Here, $\mathcal{N}$ is the normal distribution and ${\rm erf}$ is the error function. When $\eta_{\rm eff}=0$, this distribution reduces to the regular normal distribution, $ \mathcal{N} \left( \chi_{\rm eff} | \mu_{\rm eff}, \sigma_{\rm eff}^2 \right)$. A positive $\eta_{\rm eff}$ gives a distribution with a broader tail on the right, whereas a negative $\eta_{\rm eff}$ yields a distribution with a broader tail on the left. While the \textsc{skewnormal} distribution has support everywhere in $\mathbb{R}$, here we truncate it in $[-1, 1]$ to capture the physical range of $\chi_{\rm eff}$. This model was also used by \cite{Adamcewicz:2023mov} to test for correlations between $\chi_{\rm eff}$ and $q$.

To account for model-dependent conclusions, we introduce a second model that also allows for skewness but parameterizes it differently. The $\epsilon$-\textsc{skewnormal} model~\citep{Mudholkar:2001} stitches two normal distributions with the same location parameter at the mode in a way that is continuous and differentiable, i.e.
\begin{equation}
\small
 p(\chi_{\rm eff} | \mu_{\rm eff}, \sigma_{\rm eff}, \varepsilon_{\rm eff}) \propto
 \begin{cases}
 \mathcal{N} \left(\chi_{\rm eff} | \mu_{\rm eff}, \sigma_{\rm eff}^2 (1 + \varepsilon_{\rm eff})^2 \right)&  \chi_{\rm eff} < \mu_{\rm eff}\\
\mathcal{N} \left(\chi_{\rm eff} | \mu_{\rm eff}, \sigma_{\rm eff}^2 (1 - \varepsilon_{\rm eff})^2 \right) & \chi_{\rm eff} \geq \mu_{\rm eff}
 \end{cases} 
\end{equation}

We further truncate this to $\chi_{\rm eff} \in [-1, 1]$ as before. The $\varepsilon$ parameter produces skewness in the opposite sense to the $\eta_{\rm eff}$, i.e., a negative $\varepsilon_{\rm eff}$ gives a broader tail on the right and vice versa. At $\varepsilon_{\rm eff} = 0$, the model again reduces a regular normal distribution. Both the \textsc{skewnormal} and the $\varepsilon$-\textsc{skewnormal} models are unimodal, have three variable parameters, and have support everywhere in $\mathbb{R}$.

Finally, for comparison, we also run the \textsc{truncated normal} model~\citep{Miller:2020zox, RnP:GWTC3}, i.e.,

\begin{figure*}[ht]
 \centering
 \plottwo{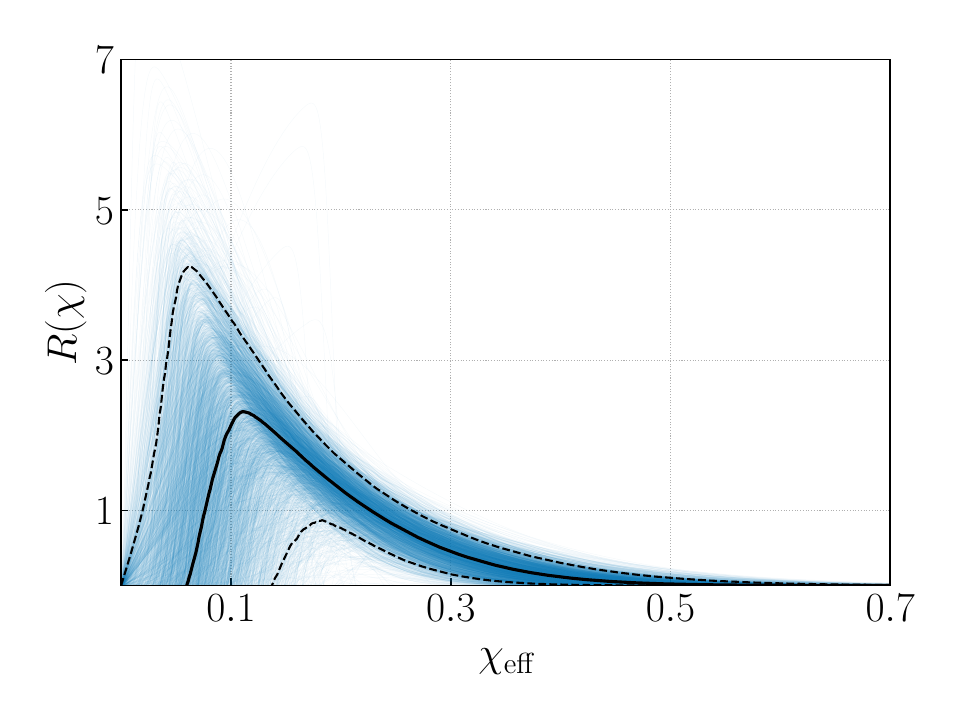}{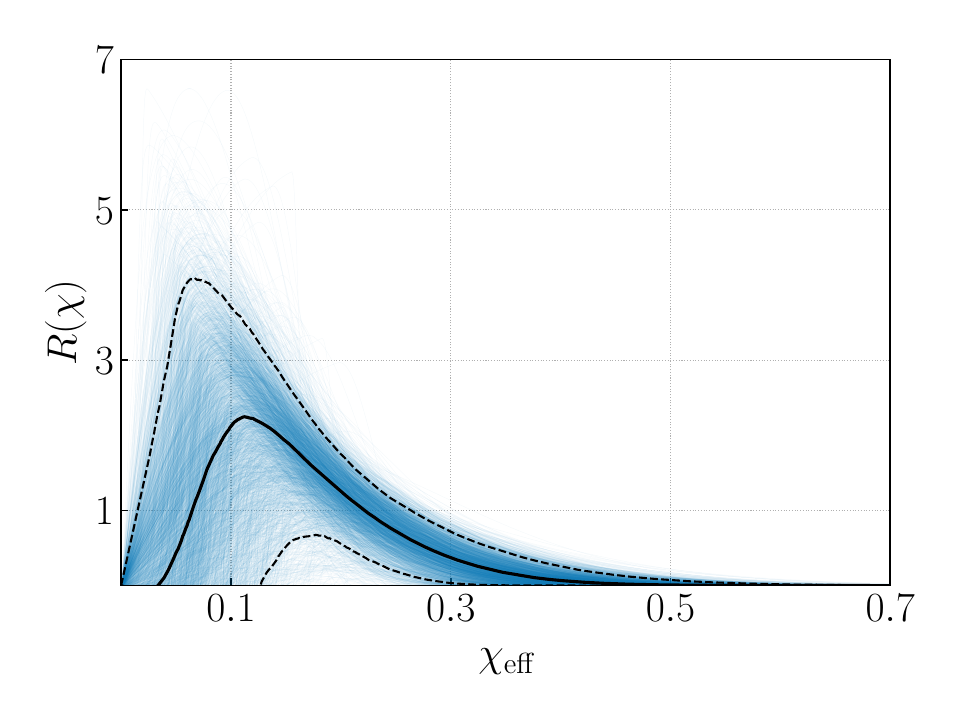}
\caption{Residual of the \textsc{skewnormal} fit on the left, and the $\varepsilon$-\textsc{skewnormal} fit on the right (as defined by Eq.~\ref{Eq:residue}). If negative effective spins are assumed to arise purely from random formation channels (like dynamical assembly in dense clusters), then $R(\chi_\mathrm{eff})$ can be interpreted as the effective spin distribution from remaining non-dynamical channels (like isolated binary evolution).
 Both show a residual spin distribution that is mostly contained at $\chi_{\rm eff} \leq 0.4$. As in Fig.~\ref{Fig:ppd}, the solid line is the median and the dashed lines are the 90\% credible levels.}
 \label{fig:residue}
\end{figure*}

\begin{equation}
 p \left( \chi_{\rm eff} | \mu_{\rm eff}, \sigma_{\rm eff}^2 \right) \propto \mathcal{N} \left( \chi_{\rm eff} | \mu_{\rm eff}, \sigma_{\rm eff}^2 \right),
\end{equation}
truncated between $\chi_{\rm eff} \in [-1, 1]$. Note that this model has only two parameters, unlike the skewnormal models.

We use these three models to hierarchically infer the \ac{BBH} candidates from the first three \ac{LVK} observing runs~\citep{Catalog:GWTC3, lvk_pe_zenodo:gwtc3}. We use uniform priors on $\mu_{\rm eff}$ and $\sigma_{\rm eff}$ for all three models. For the two skewnormal models, we also use uniform priors for $\eta_{\rm eff}$ and $\varepsilon_{\rm eff}$. In addition to $\chi_{\rm eff}$, we also model the distribution of source frame primary mass $m_1$, mass ratio $q$, redshift $z$, and the effective precessing spin $\chi_p$. The details of these distributions, the data used, and the hierarchical Bayesian formalism are covered in Appendix~\ref{Sec:Bayes}. We note in particular that all the analyses in this paper use a Gaussian model for $\chi_p$, truncated between $[0, 1]$~\footnote{Technically, the maximum value of $\chi_p$ is a function of $\chi_{\rm eff}$ and $q$~\citep{Iwaya:2024zzq}. This will impact the normalization of $p({\chi_{\rm eff}, \chi_p})$ at different parts of the parameter space. However, this effect is expected to be a small unless $|\chi_{\rm eff}| \sim 1$. Therefore, analyses in the literature generally tend to ignore this normalization, and we will do the same here.} and assume that there is no covariance between $\chi_{\rm eff}$ and $\chi_p$. {It is unlikely that this assumption will significantly impact our conclusions as there is no evidence thus far of correlations between $\chi_{\rm eff}$ and $\chi_p$~\citep{RnP:GWTC2, RnP:GWTC3}}. Table.~\ref{tab:priors} summarizes the parameters in all the analyses and the actual prior ranges used.

Figure~\ref{Fig:skewnorm_ppd} shows the $\chi_{\rm eff}$ posterior population distribution \ac{PPD} for the \textsc{skewnormal} model. Two features are immediately observable. First, the tail on the positive $\chi_{\rm eff}$ side is broader than the negative side, indicating a positive skewness. Indeed, we find that the data prefers a positive skewness with $\eta_{\rm eff} \geq 0$ at $99.1 \%$ credence and $\eta_{\rm eff} \geq 2.9$ at $90 \%$ credence. Secondly, we also find that the mode of the distribution is consistent with zero, unlike the \textsc{truncated normal} model, which shows a clear preference for a positive mode~\citep{RnP:GWTC3}. Both of these features are consistent with what we might expect from a sub-population of isolated binaries with small spins.

Figure.~\ref{Fig:eps_skewnorm_ppd} shows the \ac{PPD} for the $\varepsilon$-\textsc{skewnormal} model, which exhibits qualitatively similar features as the \textsc{skewnormal} model. This model also supports positive skewness, but the support is somewhat lower, with $\varepsilon_{\rm eff} < 0$ at $90.1\%$ credibility. With this model too, the data supports a mode that is consistent with zero. {The \textsc{skewnormal} and the $\varepsilon$-\textsc{skewnormal} models are preferred by modest Bayesfactors of 3.5 and 1.8, respectively, over the \textsc{truncated normal} model}.

Remarkably, despite being positively skewed, both the \textsc{skewnormal} and the $\varepsilon$-\textsc{skewnormal} models \textit{require} that a fraction of the population have negative $\chi_{\rm eff}$ and thereby at least one anti-aligned \ac{BH}. In fact, the lower limit on the fraction of the population with $\chi_{\rm eff} \leq 0$ remains consistent at $\sim 20\%$ between both the skewed and the \textsc{truncated normal} models, suggesting that this is a robust requirement of the data. The inferred presence of binaries with negative effective spin also suggests a potential alternative interpretation of the skewed $\chi_\mathrm{eff}$ distributions as a superposition between random and preferentially-aligned spin channels, with the former responsible for negative effective spins and zero mode and the latter driving the skewness towards positive $\chi_\mathrm{eff}$. Table~\ref{tab:skewnormal_results} shows the breakdown of $p(\chi_{\rm eff} < 0)$ for all three models.

\begin{table*}[ht]
\centering
\input{skewnormal_table}
\caption{Asymmetry metrics of the $\chi_{\rm eff}$ distribution as defined by Eq.~\ref{Eq:integrated_asymmetry} for the three unimodal models tested in Sec.~\ref{Sec:skew_eff_spin}. {Here, $\alpha (\chi_{\rm mode})$ is the asymmetry about the mode of the distribution as defined by Eq. \ref{Eq:integrated_asymmetry}.}}
\label{tab:skewnormal_results}
\end{table*}

\subsection{Measures of Asymmetry}

To draw astrophysical conclusions from these models, it is useful to define some metrics of asymmetry. Let us define one such generic measure $\alpha(\chi_0)$ about some point $\chi_0$ to be the difference between integrated probabilities of the $\chi_{\rm eff}$ distribution above and below $\chi_0$, i.e.
\begin{equation}
\small
 \alpha (\chi_0) = \int_{\chi_{\rm eff} > \chi_0} d \chi_{\rm eff} p (\chi_{\rm eff}) \, - \int_{\chi_{\rm eff} < \chi_0} d \chi_{\rm eff} p (\chi_{\rm eff}).
\label{Eq:integrated_asymmetry}
\end{equation}
For the three models considered thus far, Table~\ref{tab:skewnormal_results} shows the measured asymmetries about $\chi_\mathrm{eff}=0$, as well as the asymmetries measured about the inferred mode of the $\chi_\mathrm{eff}$ distribution. We focus in particular on $\alpha (0)$ since we expect a subpopulation from a random spin channel to be completely symmetric about $\chi_{\rm eff} = 0$. This implies that any asymmetry about it can be considered to be entirely caused by a preferentially aligned population. As seen in Tab.~\ref{tab:skewnormal_results}, there is an asymmetry of at least $12 \% - 17 \%$ about zero among the various models.

If we assume that the preferentially-aligned binaries are from isolated binary evolution and that the channel cannot produce binaries with anti-aligned spins, i.e., $\chi_{\rm eff} \geq 0$ for this sub-population, then $\alpha(0)$ would give exactly the fraction of the aligned population. This then offers an avenue for isolating and investigating the spin distribution of binaries forming from preferentially-aligned spin channels.

However, more conservatively, if we assume that preferentially aligned binaries can but are not required to have a positive $\chi_{\rm eff}$ (allowing for e.g., strong natal kicks or internal gravity-wave spin-up~\citep{Baibhav:2024}), then $\alpha(0)$ can instead be treated as a conservative lower limit on the fraction of this population. Therefore, we can say that at least $12\% - 17\%$ of the population must come from a preferentially aligned formation channel. Since random spin channels yield a symmetric $\chi_\mathrm{eff}$ distribution about zero, we can subtract this symmetric distribution to obtain the residual excess of systems arising from preferentially-aligned spin channels with $\chi_\mathrm{eff}>0$.
Let us define this residual distribution as $\mathcal{R}(\chi)$,
\begin{equation}
R (\chi) = p (\chi_{\rm eff} = \chi) - p (\chi_{\rm eff} = -\chi), \quad \chi \geq 0
 \label{Eq:residue}
\end{equation}

Figure~\ref{fig:residue} shows the residue for the \textsc{skewnormal} and the $\varepsilon$-\textsc{skewnormal} models. Any excess at positive $\chi_{\rm eff}$ values is mostly contained at $\chi_{\rm eff} \leq 0.4$. $R (\chi)$ provides information about the structure of the spin distribution in the preferentially-aligned spin channel. It suggests that the spins from such a channel are mostly small, contained within $\chi \lesssim 0.2$ with a tail extending up to $0.4$.

\begin{figure*}[t]
\centering
\includegraphics[width=0.85\textwidth]{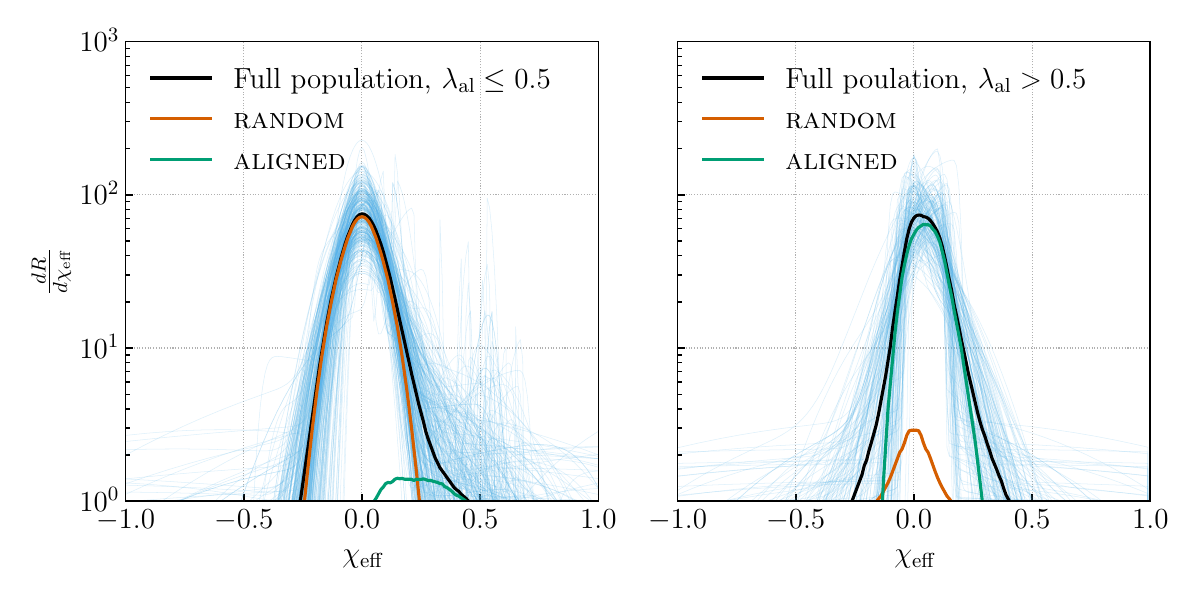}
\caption{PPDs of the two modes of $\lambda_{\rm al}$ in the mixture model analysis of Sec.~\ref{Sec:mixture_models}. The plot on the left shows the $\chi_{\rm eff}$ distribution of the $\lambda_{\rm al} \leq 0.5$ mode, while the plot on the right shows the $\lambda_{\rm al} > 0.5$ mode. The orange and green solid lines show the medians of the random and aligned populations within each mode, while the solid black line shows the median of the total population.}
\label{fig:skewnorm_modal_plots}
\end{figure*}

\section{Bimodality}
\label{Sec:mixture_models}

In the previous section, we analyzed the $\chi_{\rm eff}$ distribution for signs of asymmetry and skewness using unimodal models. We now turn towards tests of multimodality and general non-Gaussian nature. For simplicity, we restrict ourselves to the case with at most two separate populations in the data: a random and a preferentially-aligned spin channel. We model the random spin channel using a \textsc{truncated normal} distribution centered at zero. To this, we add a \textsc{skewnormal} distribution representing a possible preferentially-aligned population with a branching fraction of $\lambda_{\rm al}$. This captures our intuition that the random spin channel must be symmetric about zero, while the preferentially-aligned spin channel can have a flexible mode and a distribution that is skewed about it. Therefore, the total population distribution is given by the mixture model,

\begin{equation}
\begin{split}
p(\chi_{\rm eff}) = (1 - \lambda_{\rm al}) \, & \mathcal{N}( \chi_{\rm eff} | 0, \sigma_{\rm r}^2) \, + \\ & \lambda_{\rm al} \, {\rm SN} ( \chi_{\rm eff} | \mu_{\rm al}, \sigma_{\rm al}^2, \eta_{\rm al}),
\end{split}
\label{Eq:sn_mixture_model}
\end{equation}
with ${\rm SN} ( \chi_{\rm eff} | \mu_{\rm al}, \sigma_{\rm al}^2, \eta_{\rm al})$ defined as in Eq.~\ref{Eq:skewnormal}. Here, $\mu_{\rm al}$, $\sigma_{\rm al}$, and $\eta_{\rm al}$ are the location, scale, and skewness parameters for the \textsc{skewnormal} model, while $\sigma_{\rm r}$ is the scale of the \textsc{truncated normal} model.

We once again employ hierarchical modeling using data from the first three observing runs, with the same mass, redshift, and $\chi_p$ models as before. Figure~\ref{fig:skewnorm_mixture_model} shows a corner plot of the spin parameters of this analysis. We do not find any significant evidence for clearly distinct populations. Specifically, a near-equal mixture of aligned and random populations is disfavored, albeit not ruled out. Consequently, we fail to find any strong signs of bimodality in the $\chi_{\rm eff}$ space. This is also consistent with the results of \cite{Callister:2022qwb} which models the $\chi_{\rm eff}$ distribution as a mixture-model sum of two independent Gaussian distributions but does not observe any signs of bimodality.

Interestingly, we find that the posterior on the inferred fraction of preferentially-aligned events, $p(\lambda_{\rm al} | d)$, is bimodal, indicating that the data currently favors one channel or the other dominating and disfavors fine-tuned comparable contributions from both channels. It is therefore useful to see what the $\chi_{\rm eff}$ distribution of the two modes looks like. In Fig.~\ref{fig:skewnorm_modal_plots}, we plot the PPD of the $\lambda_{\rm al} \leq 0.5$ and $\lambda_{\rm al} > 0.5$ modes separately. The data is consistent with two possible scenarios. The first is that mergers from random spin channels constitute a large fraction of the population and that the preferentially-aligned population is small ($\lambda_\mathrm{al}\leq 0.5$; left) with positive effective spins. If, on the other hand, the fraction of preferentially-aligned events is large ($\lambda_\mathrm{al}\geq 0.5$; right), then this aligned population itself is inferred to be centered near $\chi_\mathrm{eff} \simeq 0$.

These results qualitatively agree with the results from the unimodal models. The $\chi_{\rm eff}$ distribution is centered close to zero, but there are signs of skewness and asymmetry. In particular, the $90\%$ lower limit on the asymmetry for the low and high $\lambda_{\rm al}$ populations is $4 \%$ and $23 \%$, respectively. The lower limit on the probability mass at $\chi_{\rm eff} < 0$ is $18\%$ and $40\%$ {for the high and low modes}, respectively.

A curious feature in the low $\lambda_{\rm al}$ mode is the presence of some support for an aligned population with $\mu_{\rm al} \sim 0.5$; this can be seen both in an excess of blue traces in Fig.~\ref{fig:skewnorm_modal_plots} at $\chi_{\rm eff} \sim 0.5$, as well as in the $\mu_{\rm al} - \lambda_{\rm al}$ two-dimensional posterior in Fig.~\ref{fig:skewnorm_mixture_model}. The implication of this feature is unclear. A similar feature does not exist at $\chi_{\rm eff} \sim - 0.5$, suggesting that if this feature is real {and of hierarchical origin, it does not originate from dense star clusters. However, it is possible that hierarchical mergers from AGN disks could be the source, as some degree of preferential alignment is broadly expected from such environments~\citep{Yang:2019, Santini:2023, Cook:2024}}. While the statistical significance of this feature is not high, it will be interesting to see if it persists with future data.

\subsection{Mixture model with distinct $p(q)$}
\label{Sec:separate_pq_mixture_model}

In our analysis of multimodality thus far, we have assumed that the aligned and random populations have the same mass distributions. Now we test to see if the two population types have separate $p(q)$ distributions. This is also motivated by the correlations observed between $\chi_{\rm eff}$ and $q$~\citep{Callister:2021fpo, Adamcewicz:2022hce, RnP:GWTC3}, one possible cause of which could be the interplay between multiple channels with different spin and mass-ratio distributions.

We follow \cite{Baibhav:2022qxm}, who performed a similar analysis for spin tilts. We model the mass-ratio distribution as two independent powerlaws, $p(q) \propto q^{\beta_A}$ and $p(q) \propto q^{\beta_R}$ for the aligned and random populations, respectively. Analogous to Eq.~\ref{Eq:sn_mixture_model}, we then have

\begin{equation}
\begin{split}
p(\chi_{\rm eff}, q) \propto \, & (1 - \lambda_{\rm al}) q^{\beta_R} \, \mathcal{N}( \chi_{\rm eff} | 0, \sigma_{\rm r}^2) \, \\ + &\lambda_{\rm al} q^{\beta_A} \, {\rm SN} ( \chi_{\rm eff} | \mu_{\rm al}, \sigma_{\rm al}^2, \eta_{\rm al})
\end{split}
\end{equation}

\begin{figure}[t]
\centering
\includegraphics[width=0.52\textwidth]{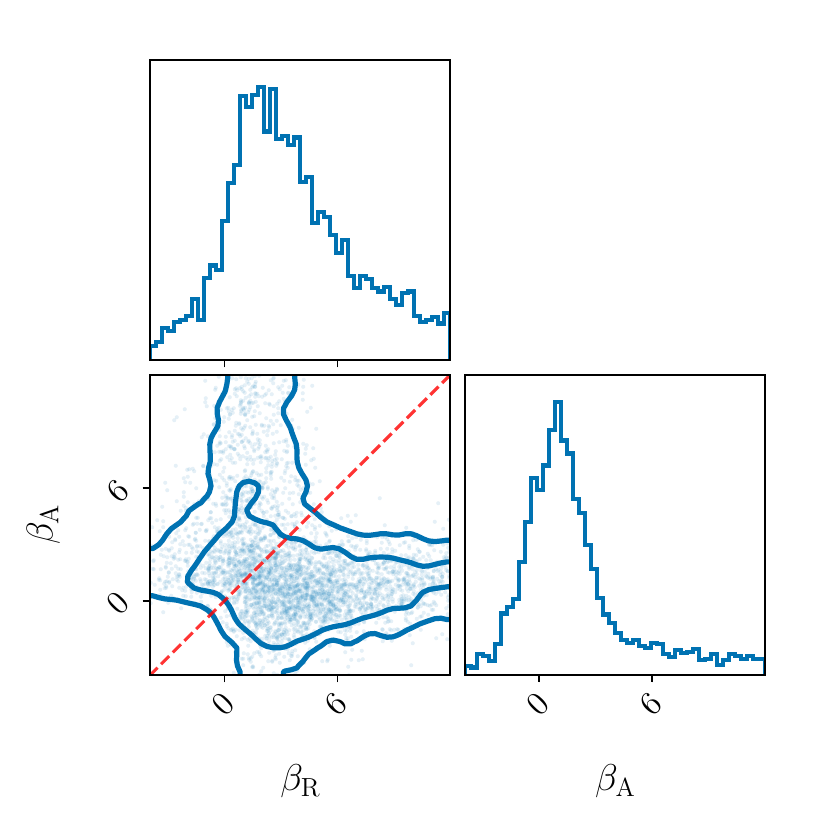}
\caption{The posterior distributions of the mass-ratio power-law indices for the aligned population ($\beta_A$) and the random population ($\beta_R$). We find that the data has a mild preference for $\beta_R > \beta_A$. {The dashed red line corresponds to $\beta_R = \beta_A$. }}
\label{fig:beta_plots}
\end{figure}

Repeating the hierarchical Bayesian analysis, we find the results to be broadly consistent with those in Sec.~\ref{Sec:mixture_models} with a common mass ratio distribution. While there is no strong evidence for separate mass-ratio distributions for the two channels, the data shows a mild preference for a flatter distribution for the preferentially-aligned spin channel. Figure~\ref{fig:beta_plots} shows a corner plot of the two mass-ratio indices; we find that $\beta_{\rm R} > \beta_{\rm A}$ at $65 \%$ credence. This is broadly consistent with the findings of \cite{Baibhav:2022qxm} by modeling a mixture model of component spin tilts and mass-ratio distributions. {The distribution of all $\chi_{\rm eff}$ parameters along with $\beta_A$ and $\beta_R$ is shows in Fig.~\ref{fig:q_skewnorm_mixture_model}. The posterior distributions of the rest of the parameters are consistent with that of Fig.~\ref{fig:skewnorm_mixture_model} and the mass-ratio indices do not show any noticeable covariance with any of them.}

We comment more on the astrophysical implications of such a possibility in the next section.

\section{Discussion}
\label{Sec:discussion}

Through a series of models and tests, we have looked for signs in the $\chi_{\rm eff}$ distribution that could indicate the presence of multiple sub-populations. Let us revisit the potential features from Sec.~\ref{Sec:eff_spin_shape} in light of our results.

\begin{enumerate}
\item Skewness: Through the unimodal fits in Sec.~\ref{Sec:skew_eff_spin}, we find that there is statistical evidence for a positively skewed $\chi_{\rm eff}$ distribution. This points to an aligned subpopulation, which could exist either as a sub-dominant component of the population or as the dominant component with small spins. The results of the mixture model fit indicate that it is unlikely that both the random and preferentially-aligned spin channels constitute a similar fraction of the merging population.

\item Asymmetry about $\chi_{\rm eff} = 0$: We find that unimodal models show evidence of asymmetry about $\chi_{\rm eff} = 0$. In the \textsc{truncated normal} model, this is due to a positive location parameter. In the skewed models, the asymmetry is tied with the skewness, as the modes of these models are consistent with zero. The asymmetry allows us to place a conservative lower limit on the fraction of the aligned subpopulation at $12 \% - 17\%$. This estimate makes no particular assumption about the shape of $p(\chi_{\rm eff})$ of the preferentially-aligned spin channels or their ability (or inability) to make binaries with negative $\chi_{\rm eff}$.

\item Multimodality: In Sec.~\ref{Sec:mixture_models}, we tested for signs of bimodality using a mixture model accounting for a random and an aligned population. We find no strong evidence for bimodality; in fact, the data favors either a low or high $\lambda_{\rm al}$. We also do not find strong evidence that the aligned and the random sub-populations have separate $p(q)$ distributions.

\end{enumerate}

All in all, these results indicate the existence of an aligned population whose spins are probably small. They also indicate that { at least about $\sim 20\%$ of the population} has negative spins, consistently across all three unimodal models. If one assumes that preferentially-aligned spin pathways such as isolated binary evolution cannot produce systems with negative $\chi_{\rm eff}$, our results suggest that {at least about $40\%$ of merging \acp{BBH}} come from some sort of dynamical formation channel~\footnote{The doubling is due to the the symmetry of such formation channels about $\chi_{\rm eff} = 0$. }.

\subsection{$\chi_{\rm eff} - q$ correlations}
\label{sec:chi_eff_q_corr}

Next, we consider the relation between the skewness we see here and the $\chi_{\rm eff} - q$ anti-correlation observed in \ac{GW} data~\citep{Callister:2021fpo, Adamcewicz:2022hce, Catalog:GWTC3}. In fact, the marginal $\chi_{\rm eff}$ distribution accounting for the correlation between $\chi_{\rm eff}$ and $q$ looks quite similar to what we observe with our skewed fits -- see Fig. 12 of \cite{Callister:2021fpo} and Fig. 4 of \cite{Adamcewicz:2022hce}.

This is not very surprising; the observed correlation between $\chi_{\rm eff}$ and $q$ suggests that there is a subset of the binaries that have preferentially positive spins, a behavior that random spin channels cannot easily produce. Therefore, the correlation itself is an indicator of the presence of preferentially-aligned spin channels. Similarly, skewness in the $\chi_{\rm eff}$ marginal distribution is a sign of the presence of a preferentially-aligned spin channel.

Another natural question is if the $\chi_{\rm eff} - q$ correlation primarily arises from the physics of a single preferentially-aligned spin channel or the interplay between different channels with different spin and mass-ratio distributions. Intriguingly, in our mixture model analysis in \ref{Sec:separate_pq_mixture_model}, we find that there is some mild support for $\beta_{R} \geq \beta_{A}$. If this is indeed true, i.e., if the \ac{BBH} population consists of a random formation channel that has smaller $\chi_{\rm eff}$ values and larger $q$ values and an aligned population with lower $q$ values and higher $\chi_{\rm eff}$ values, it would be qualitatively consistent with the correlation between the parameters. This is a tantalizing possibility and something that future data from O4 and beyond might be able to shed light on.

\subsection{ {Is there a sharp excess} of non-spinning \acp{BBH}?}

There has been debate in the recent literature over the existence of \acp{BBH} that are clearly precessing or have anti-aligned spins. Signs of precession have been reported in certain individual systems such as GW190521 and GW200129\_065458~\citep[e.g.][]{Catalog:GWTC3, Hannam:2021pit, LIGOScientific:2020ufj, Miller:2023ncs, Hoy:2024qpy}. Another candidate from O3, GW191109\_010717, has been reported to have negative effective spins~\citep{Catalog:GWTC3, Islam:2023zzj}. However, such evidence for precession or anti-aligned spins is not incontrovertible. Data quality issues at LIGO Livingston at the times of GW200129\_065458 and GW191109\_010717 could be impacting our estimation of spins~\citep{Payne:2022spz, Udall2024}. On the other hand, evidence for precession in GW190521 is found only when the merger and the last pre-merger cycle are analyzed together~\citep{Miller:2023ncs}, and the small number of detected cycles leaves it particularly open to other interpretations such as eccentric mergers or hyperbolic encounters~\citep{Romero-Shaw:2020thy, Gayathri:2020coq, Gamba:2021gap}.

\begin{figure}[t]

 \centering

 \includegraphics[width=0.52 \textwidth]{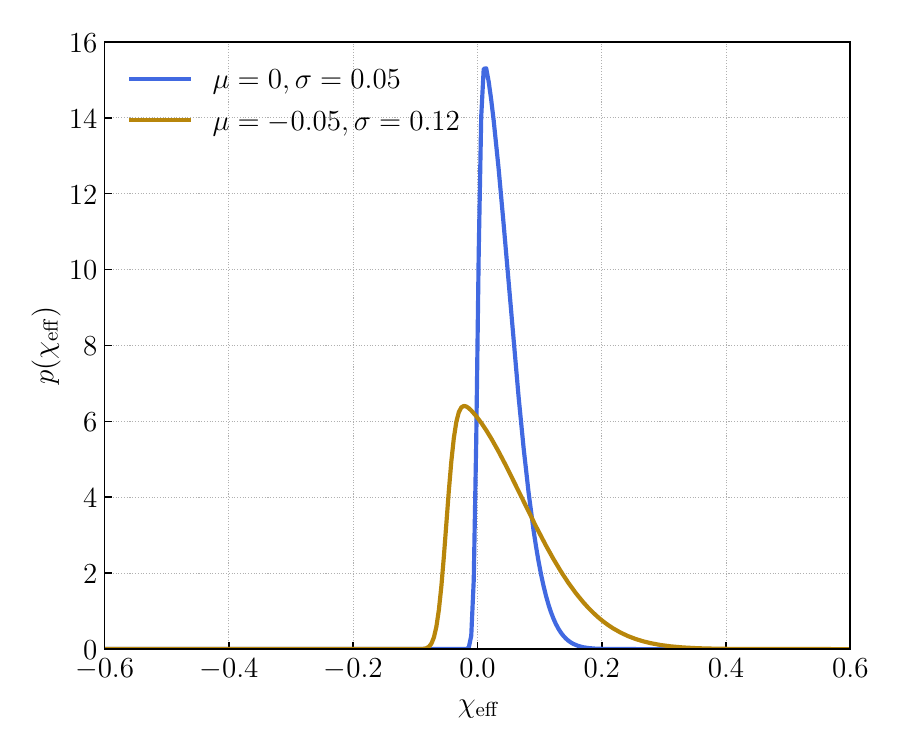}

\caption{{Two different instances of the \textsc{skewnormal} model. The blue curve shows an instance with high skewness ($\eta = 10$) but with no probability mass at $\chi_{\rm eff} < 0$. The gold-colored curve, on the other hand, shows a distribution with the same skewness but with significant probability mass at $\chi_{\rm eff} < 0$. The fact that the analysis described in Sec.~\ref{Sec:skew_eff_spin} find almost no support for the former but significant support for the latter illustrates the informative power of the skewnormal models regarding the question of anti-aligned spins}}

\label{fig:skewnormal_illustrative}

\end{figure}

There is also a debate at the population level. Using the \textsc{truncated normal} fit to data from the first two observing runs, \cite{Miller:2020zox} found that a significant fraction of \ac{BBH} mergers should have negative effective spins, indicating that at least one spin is anti-aligned with respect to the orbital angular momentum. This was supported by \cite{RnP:GWTC2} using data until the first half of the O3 run; moreover, they also found evidence for anti-aligned spins when modeling the population distribution of component spin tilts.

However, \cite{Roulet:2021hcu} fit the $\chi_{\rm eff}$ distribution with a three-component mixture model that explicitly includes a non-spinning population and found that the data favors the interpretation that there are systems with positive $\chi_{\rm eff}$ and a excess of systems with small spins. The fraction of anti-aligned systems (with negative $\chi_{\rm eff}$) was consistent with zero in their analysis. \cite{Galaudage:2021rkt} reached a similar conclusion using models for the component-spin distribution that allow for a non-spinning sub-population. \cite{Tong:2022iws} updated these results to include data from GWTC-3, and found less clear evidence for {a sharp excess} with small spins, concluding that the data is yet not informative enough to resolve a spike-like feature with certainty. \cite{Callister:2022qwb} argued that a degeneracy between the `spike' at zero spin and the mean of the bulk population causes the former to be overestimated. They further show through a set of null tests that similar statistical significance for the spike as found in the real data can also be found in simulated catalogs without {a sharp excess} of non-spinning \acp{BBH}.
Subsequent analyses, including measurements with more complex parametric~\citep{Vitale:2022dpa} and non-parametric models~\citep{Edelman2023,Callister:2023tgi,Heinzel2024}, infer the existence of events with anti-aligned spins.

This question -- do the data support {systems with anti-aligned spins or a sharp excess of non-spinning systems} -- is of significant astrophysical importance in understanding the formation channels of \acp{BBH} and the physics of massive stellar evolution. The absence of anti-aligned systems would suggest that most merging binaries come from the field formation scenario, with the excess of non-spinning systems construed as evidence for efficient angular momentum transport in massive stars~\citep{Spruit:2001tz, Fuller:2019sxi}. Conversely, the absence of a non-spinning excess can be interpreted as evidence that the physics of angular momentum transport is more complex and varied. Incontrovertible evidence for anti-aligned spins, especially with large magnitudes, can be interpreted as evidence for mergers from the various cluster formation scenarios~\citep{Rodriguez:2016vmx}.

The skewnormal models used in Sec.~\ref{Sec:skew_eff_spin} can inform some of these debates. These models have two strengths that make them useful for this purpose: first, since the two tails of the {model} can be independently varied, they have the flexibility to fit either populations that have anti-aligned binaries and and populations with an excess $\chi_{\rm eff}=0$. For instance, At the same time, they avoid some of the degeneracies that are inherent in mixture models that can make it harder to draw conclusions about spins, as pointed out by \cite{Callister:2022qwb}. As described in Sec.~\ref{Sec:skew_eff_spin}, while we find that most systems have small effective spins -- the mode of the skewnormal fits is consistent with zero -- our models do not find {a large excess} of systems there and require that systems with negative $\chi_{\rm eff}$ exist. 

We find that at least $\sim20\%$ of the population in our skewnormal fits have negative effective spins, an estimate consistent with the \textsc{truncated normal} model. Assuming that it is hard for preferentially-aligned spin channels to form binaries with negative $\chi_{\rm eff}$, this suggests that at least $\sim 40\%$ of the systems should indeed arise from a random formation channel such as clusters. {These limits are broadly consistent with analyses that make stronger physics-inspired modeling choices~\citep{Zevin:2020gbd, Wong:2020ise}. However, care is needed in such comparisons as the underlying population synthesis models can be subject to numerous theoretical uncertainties. Such studies in the literature generally conclude that neither dynamical or isolated binary evolution can explain the entirety of the observed \acrodef{bbh} population and our results agree with that. Finally,} as can be seen from Fig.~\ref{Fig:skewnorm_ppd}, the magnitude of these negative spins are small, limited to $\chi_{\rm eff} \geq -0.3$, which also bolsters the limits placed on the hierarchical merger products by \cite{Fishbach:2022lzq}.

A note of caution is warranted on the interpretation here. The skewnormal models have the ability to fit a distribution with an excess of slow spinning systems. However, if most systems have very tiny spins~\citep{Fuller:2019sxi}, they might struggle to fit such a delta-function like distribution owing to a lack of resolving power with current data. However, in such a case, since the skewnormal models have flexible tails, one would expect that our posteriors would also show some support with a vanishing negative tail. That this is not the case {(see Fig.~\ref{fig:skewnormal_illustrative} for an illustrative plot of what such a  looks like)} lends some credence that the data does not support such a sharp spike. 

{To further investigate this, we compared a mixture model comprising a non-spinning delta function spike and the \textsc{skewnormal} model (similar to \cite{Tong:2022iws, Adamcewicz:2023szp}) and found that it is disfavored compared to the standalone \textsc{skewnormal} with a $\log \mathcal{B} = -3.0$. However, while investigating this feature, we discovered that previous analyses modeling a peak at zero spin ~\citep{Tong:2022iws, Adamcewicz:2023szp}) suffered from a normalization error in the population likelihood that led to a slight overestimate of the evidence for a non-spinning subpopulation. While this potentially brings it to closer alignment with \cite{Callister:2022qwb}, it difficult to compare this number for the skewnormal models directly to the results of these previous works. While the qualitative conclusions of \cite{Tong:2022iws} and \cite{Adamcewicz:2023szp} appear unchanged, a more thorough investigation is beyond the scope of this paper and will be revisited in \cite{Adamcewicz:2025}.}

\section{Conclusion}

In this paper, we have analyzed the shape of the $\chi_{\rm eff}$ distribution of \ac{BBH} mergers to look for signatures of a multiplicity of formation channels. We found that the distribution shows robust signs of skewness and asymmetry when empirical models that allow for those things are used. We do not find strong evidence for bimodality. These results suggest the existence of both a preferentially-aligned subpopulation, probably with small but positive spins, and a subpopulation with $\chi_{\rm eff} \leq 0$.

There are several directions for future inquiry this work suggests. It would be interesting to see if any of the unimodal skewnormal models show further support for correlations with other parameters. One form of this was already tested in \cite{Adamcewicz:2023mov} where they still find a $\chi_{\rm eff} - q$ anti correlation with the \textsc{skewnormal} model. Since skewness and asymmetry can be explicitly identified as arising from aligned sub-populations, this could allow us to see what fraction of the correlation is true.

In Sec.~\ref{Sec:separate_pq_mixture_model}, we tested for the possibility that the two components in our mixture model also have two different mass-ratio distributions and found that there is not enough evidence to support that. \cite{Ray:2024hos} and \cite{Antonini2024} have recently found that it is possible that different features in the mass spectrum might correspond to populations with different spin distributions. If real, this could point to a cleaner separation of the populations. More data could allow a mixture-model analysis with our skewnormal models to find stronger evidence for such separations.

\section*{Acknowledgements}
We are grateful to Sylvia Biscoveanu and Eric Thrane for helpful discussion and suggestions. We further thank Simona Miller for comments on the manuscript. S.B. acknowledges support from NSF grant PHY-2207945 (PI: Kalogera) and from Australian Research Council (ARC) grant CE230100016. T.~C.~ is supported by the Eric and Wendy Schmidt AI in Science Postdoctoral Fellowship, a Schmidt Sciences program. Z.D. was supported through the CIERA Board of Visitors Research Professorship. V.K. was supported by the Gordon and Betty Moore Foundation (grant awards GBMF8477 and GBMF12341), through a Guggenheim Fellowship, and the D.I. Linzer Distinguished University Professorship fund. The authors are grateful for computational resources provided by the LIGO Laboratory and supported by NSF Grants No. PHY-0757058 and No. PHY-0823459. This material is based upon work supported by NSF's LIGO Laboratory which is a major facility fully funded by the National Science Foundation. This research has made use of data obtained from the Gravitational Wave Open Science Center (\href{https://gwosc.org}{gwosc.org}), a service of LIGO Laboratory, the LIGO Scientific Collaboration, the Virgo Collaboration, and KAGRA. This document carries a LIGO DCC number P2400571.

\bibliography{ast_paper}{}
\bibliographystyle{aasjournal}

\appendix

\section{Hierarchical Bayesian Inference}
\label{Sec:Bayes}

We perform hierarchical Bayesian inference~\citep{Thrane:2019, Vitale:2020aaz} using \ac{GW} data from the first three observing runs (O1 + O2 + O3) of the \ac{LVK} observing runs~\citep{Catalog:GWTC1, Catalog:GWTC2, Catalog:GWTC3, lvk_pe_zenodo:gwtc3}. {We follow \cite{RnP:GWTC3} for selecting the list of detections to include in our analysis. We impose a false alarm rate threshold of $1 \text{yr}^{-1}$ giving us a total of 69 events. GW190814 is left out because of the ambiguous nature of its secondary. Hierarchical Bayesian inference involves calculating hyperparameters $\Lambda$ that describe the population distribution}

\begin{equation}
    p(\Lambda | \{ d\}) \propto \pi(\Lambda) \prod_i  \frac{\int d \theta \mathcal{L}(d_i | \theta) \pi(\theta | \Lambda)}{\zeta(\Lambda)}
\end{equation}
{where $ \pi(\Lambda)$ is the prior on the hyperparameters, $\mathcal{L}(d_i | \theta)$ is the \ac{GW} likelihood and $i$ is an index over detected events. The denominator $\zeta(\Lambda)$ accounts for selection effects and is the fraction of detected events (i.e. with false alarm rate of less than $1 \text{yr}^{-1}$) for the astrophysical population distribution given by $\pi (\theta | \Lambda)$~\citep{Mandel:2018mve, Vitale:2020aaz}. To calculate $\zeta(\Lambda)$ we use the simulated signal sets released by the \ac{LVK} at the end of the O3 run~\citep{lvk_senitivities_zenodo:gwtc3}}.

We model the primary source-frame mass distribution as a mixture model of a power law and a Gaussian at higher masses~\citep{Talbot:2018cva}, {sometimes referred to as \textsc{powerlaw+peak} in the literature~\citep{RnP:GWTC2, RnP:GWTC3}. The full distribution on the primary mass is given by}

\begin{equation}
 \begin{split}
        p(m_1 | \alpha, \lambda_{\rm peak}, \mu_{\rm peak}, \sigma_{\rm peak}, m_{\rm min}, m_{\rm max}, \delta_m) \, = \, &  \Big[\lambda_{\rm peak} \mathcal{N} (m_1 | \mu_{\rm peak}, \sigma_{\rm peak}) \, + \\ & (1 - \lambda_{\rm peak}) \, m_1^{-\alpha} \Big]  S(m_1 | m_{\rm min}, \delta_m) \, \Theta(m_1 \geq m_{\rm min}) \, \Theta(m_1 \leq m_{\rm max}),
 \end{split}
\end{equation}
{where $\Theta$ is the Heaviside function and $S(m_1 | m_{\rm min}, \delta_m)$ is a flexible smoothing function that tapers the distribution to zero at low mass, with $\delta_m$ being the width of the taper. The secondary mass is indirectly modeled using a power law in $q$, again with a low-mass taper in $m_2$}:
\begin{equation}
 p(q | \beta_q) \propto q^{\beta_q} S (q m_1 | m_{\rm min}, \delta_m)
\end{equation}
{We direct the reader to the appendices of \cite{RnP:GWTC2, RnP:GWTC3} for a detailed description of the mass model.} 

We also use a powerlaw for redshift $z$ to allow for evolving merger rates with redshift~\citep{Fishbach:2018edt}, i.e.,
\begin{equation}
p(z | \kappa_z) \propto (1 + z)^{\kappa_z}
\end{equation}

In addition to the effective inspiral spin, for all models, we also fit the effective precessing spin $\chi_p$, a variable that captures in-plane spins~\citep{Hannam:2013oca, Schmidt:2014iyl}

\begin{equation}
 \chi_p = {\rm max} \left \{ \chi_1 \sin \theta_1, \left( \frac{4q + 3}{3q + 4}\right) q \chi_2 \sin \theta_2\right \}
\end{equation}

We fit $\chi_p$ using a normal distribution truncated between $[0, 1]$.

\begin{equation}
\small
 p(\chi_p| \mu_{\rm p}, \sigma_{\rm p}) \propto
 \begin{cases}
 \mathcal{N} \left(\chi_{\rm p} | \mu_{\rm p}, \sigma_{\rm p}^2 \right)& 0 \leq \chi_p < 1\\
0 & {\rm otherwise}
 \end{cases} 
\end{equation}

Our modeling of $\chi_p$ has a minor difference to the analysis in \cite{RnP:GWTC3} in that they also allow for covariance between $\chi_{\rm eff}$ and $\chi_p$. However, they find that the data is broadly uninformative and the covariance between $\chi_{\rm eff}$ and $\chi_p$ is consistent with zero. Therefore, we assume a priori that the two spin parameters are uncorrelated. {Tab.~\ref{tab:priors} describes the parameters being used in all our analyses and their prior ranges.}

In all our analyses, we control errors in Monte Carlo averages by cutting posterior samples with large estimated variances in their log-likelihoods~\citep{Essick:2022ojx, Talbot:2023pex} rather than on $n_{\rm eff}$. Specifically, we adopt a threshold of 1 for the variance of the log of the estimator. {Finally, we implement our analysis using the \textsc{GWpopulation}~\citep{Talbot:2019} and \textsc{bilby}~\citep{Ashton:2019} packages, using the nested sampler \textsc{dynesty}~\citep{Speagle:2020} and \textsc{jax} for GPU-based Bayesian inference.}  

\begin{table*}[h]
\centering
 \input{priors_table}
 \caption{Parameters and priors for the various models}
\label{tab:priors}
\end{table*}

\newpage
\section{Corner Plots}

\begin{figure}[h]
 \centering
 \includegraphics[width=1.05\textwidth]{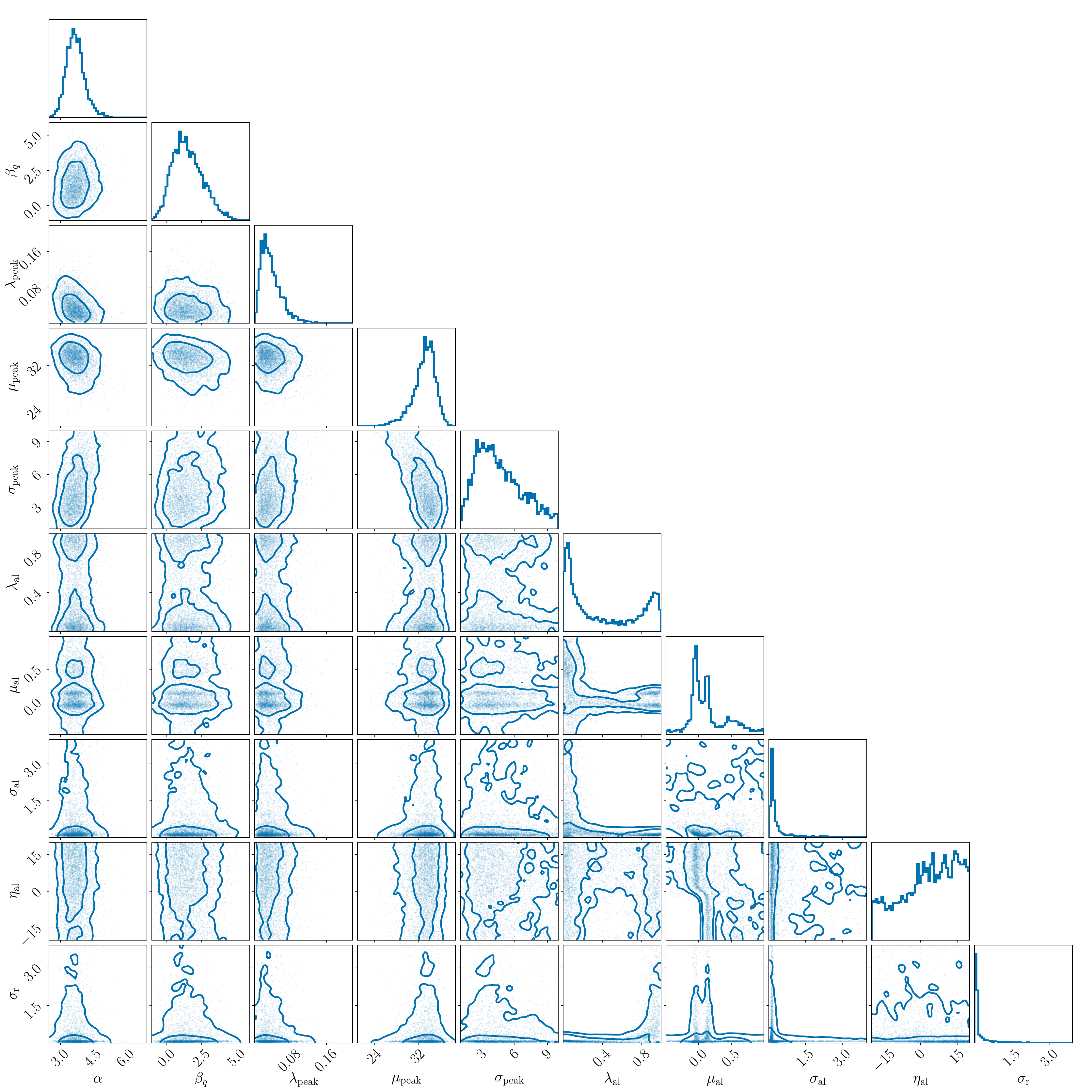}
 \caption{Corner plot for the parameters of {the $m_1$ and} $\chi_{\rm eff}$ distribution for the skewnormal mixture model. Note in particular the distribution of $\lambda_{\rm al}$ which is bimodal. This suggests that the population is either dominated by either a random spin channel or a preferentially-aligned spin channel. A comparable mixture of both channels is disfavored. {No covariance is evident between the mass and spin hyperparameters}.}
 \label{fig:skewnorm_mixture_model}
\end{figure}

\begin{figure}[h]
 \centering
 \includegraphics[width=0.95\textwidth]{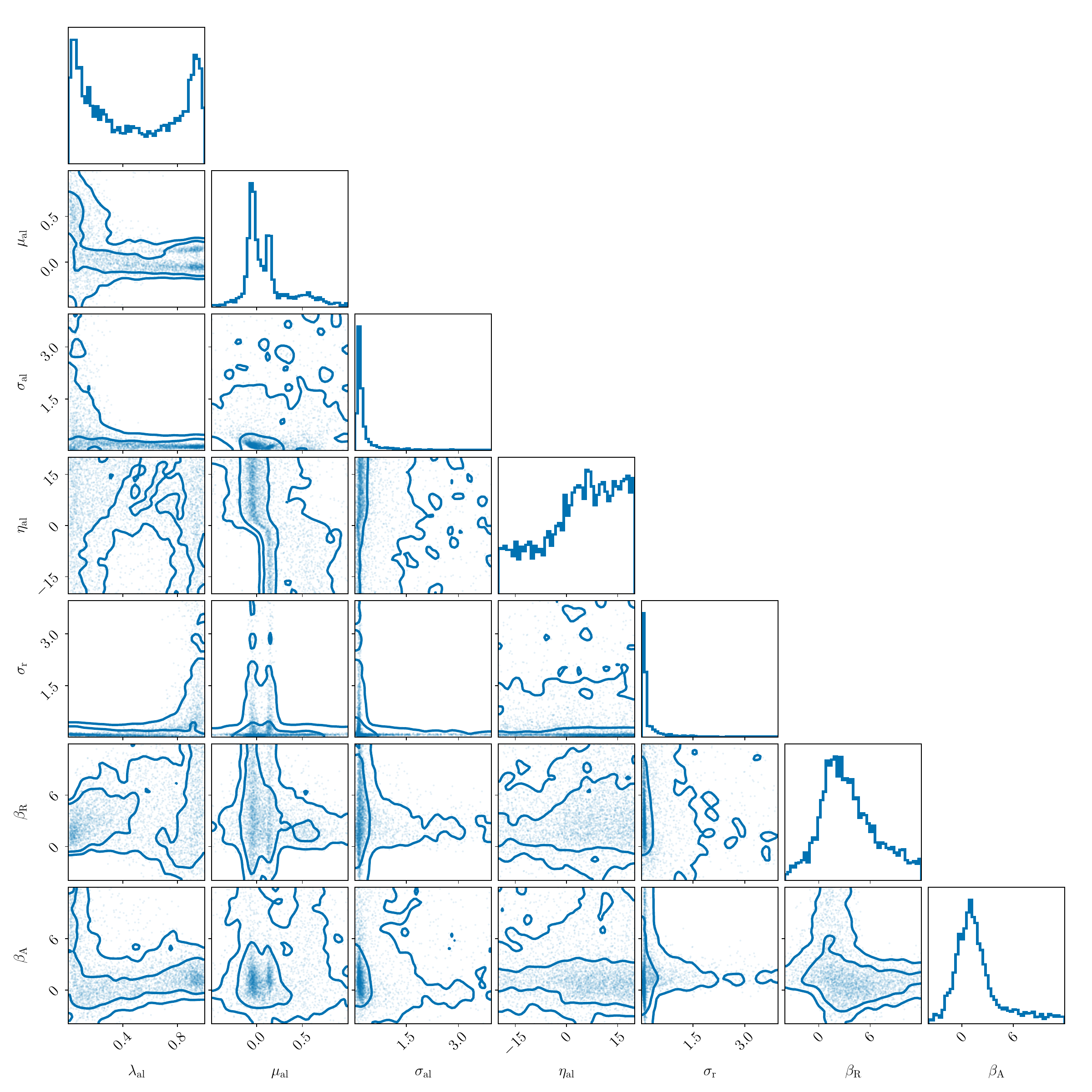}
 \caption{{Corner plot for the parameters of the $\chi_{\rm eff}$ distribution for the mixture model with distinct $p(q)$ described in Sec.~\ref{Sec:separate_pq_mixture_model}. We see the same bimodality in $\lambda_{\rm al}$ as in Fig.~\ref{fig:skewnorm_mixture_model}. The power-law indices $\beta_A \text{ and } \beta_R$ are not strongly correlated with any other parameters. The only new structure seen is that, as expected, $\beta_R (\beta_A$) becoming unconstrained at $\lambda_{\rm al} = 1 (\lambda_{\rm al} = 0 )$, or when $\mu_{\rm al} = 0$.}}
 \label{fig:q_skewnorm_mixture_model}
\end{figure}

\end{document}

%% file: skewnormal_table.tex
\begin{tabular*}{1.05\textwidth}{c  @{\extracolsep{\fill}} c c c c c } 
\hline \hline

Model & $ \alpha(0), \, 90 \%$ lower limit & $\alpha(0)$ & $ \alpha(\chi_{\rm mode}), \, 90 \%$ lower limit  &  $ \alpha (\chi_{\rm mode})$ & $p(\chi_{\rm eff} \leq 0)$,  \textbf{$90 \%$ lower limit} \\
\hline \textsc{Truncated Normal}  & $17 \%$  & $0.37^{+0.25}_{-0.26}$ & -- & --& $21\%$ \\ 
\textsc{Skewnormal} &  $12\%$ & $0.34^{+0.29}_{-0.28}$ & $24\%$ & $0.63^{+0.15}_{-0.50}$ &$22\%$\\
$\varepsilon-$\textsc{Skewnormal} & $13 \%$ & $0.35^{+0.27}_{-0.29}$ & $1\%$& $0.50^{+0.45}_{-0.65}$ &$22\%$\\ 
\hline
\end{tabular*}

%% file: priors_table.tex
\begin{tabular*}{0.95\textwidth}{c  @{\extracolsep{\fill}} c c } 
\hline \hline
Parameter & Description & Prior \\
\hline 
\multicolumn{1}{c}{}  & \multicolumn{1}{c}{Commmon priors} \\
\hline 
$\kappa_z$ & Power-law index of the redshift distribution & U$(-6, 6)$ \\
$\alpha$ & Power-law index of $m_1$ in the \textsc{powerlaw+peak} model   & U$(-4, 12)$ \\
$\beta_q$ & Power-law index of $q$ & U$(-4, 12)$ \\
$m_{\rm min}$ & Minimum mass & U$(2, 10) \, M_{\odot}$ \\
$m_{\rm max}$ & Maximum mass & U$(30, 100) \, M_{\odot}$ \\
$\delta_m$ & Taper size at low mass end & U$(0.01, 10) \, M_{\odot}$ \\
$\lambda_{\rm peak}$ & Fraction of \acp{BBH} from the Gaussian peak of the \textsc{powerlaw+peak} model & U$(0, 1)$\\
$\mu_{\rm peak}$ & Location of the Gaussian peak in the \textsc{powerlaw+peak} model & U$(20, 50) \, M_{\odot}$ \\
$\sigma_{\rm peak}$ & Width of the Gaussian peak in the \textsc{powerlaw+peak} model & U$(1, 10) \, M_{\odot}$ \\
\hline
\multicolumn{1}{c}{}  & \multicolumn{1}{c}
{\textsc{skewnormal} model} \\
\hline  
$\mu_{\rm eff}$ & Location parameter of the $\chi_{\rm eff}$ distribution & U$(-1, 1)$ \\
$\sigma_{\rm eff}$ & Scale parameter of the $\chi_{\rm eff}$ distribution & U$(0.01, 4)$ \\
$\eta_{\rm eff}$ & Skewness parameter of the $\chi_{\rm eff}$ distribution & U$(-20, 20)$ \\
\hline
\multicolumn{1}{c}{}  & \multicolumn{1}{c}{$\varepsilon$-\textsc{skewnormal} model} \\
\hline  
$\mu_{\rm eff}$ & Location parameter of the $\chi_{\rm eff}$ distribution & U$(-1, 1)$ \\
$\sigma_{\rm eff}$ & Scale parameter of the $\chi_{\rm eff}$ distribution & U$(0.01, 4)$ \\
$\varepsilon_{\rm eff}$ & Skewness parameter of the $\chi_{\rm eff}$ distribution & U$(-1, 1)$ \\
\hline
\multicolumn{1}{c}{}  & \multicolumn{1}{c}{\textsc{truncated normal} model} \\
\hline 
$\mu_{\rm eff}$ & Location parameter of the $\chi_{\rm eff}$ distribution & U$(-1, 1)$ \\
$\sigma_{\rm eff}$ & Scale parameter of the $\chi_{\rm eff}$ distribution & U$(0.01, 4)$ \\
\hline 
\multicolumn{1}{c}{}  & \multicolumn{1}{c}{\textsc{skewnormal mixture} models} \\
\hline
$\sigma_{\rm dy}$ & Scale parameter of the \textsc{random} channel & U$(0.01, 4)$ \\
$\lambda_{\rm al}$ & Fraction of the aligned population & U$(0, 1)$ \\
$\mu_{\rm al}$ & Location parameter of the \textsc{aligned} channel & U$(-1, 1)$ \\
$\sigma_{\rm al}$ & Scale parameter of the \textsc{aligned} channel & U$(0.01, 4)$ \\
$\eta_{\rm ak}$ & Skewness parameter of the \textsc{aligned} channel & U$(-20, 20)$ \\
$\beta_q^{A}$ & Mass ratio power-law index of the aligned population, if modeling independently & U$(-4, 12)$ \\
$\beta_q^{R}$ & Mass ratio power-law index of the random population, if modeling independently & U$(-4, 12)$ \\
\hline

\end{tabular*}